\documentclass[12pt]{article}
\usepackage{amsfonts,amssymb,amsmath}
\usepackage[dvips]{epsfig}
\begin{document}
\title{\bf Dephasing-assisted selective incoherent quantum transport}
\author{ Naghi Behzadi $^{a}$
\thanks{E-mail:n.behzadi@tabrizu.ac.ir}  ,
Bahram Ahansaz $^{a}$ ,
Hadi Kasani $^{a}$
\\ $^a${\small Research Institute for Fundamental Sciences,}
\\ {\small University of
Tabriz, Tabriz 51666-16471, Iran.}} \maketitle
\begin{abstract}
\noindent Selective energy transport throughout a quantum network connected to more than one reaction center can play an important role in many natural and technological considerations in photo-systems. In this work, we propose a method in which an excitation can be transported from the original site of the network to one of the reaction centers arbitrarily using independent sources of dephasing noises. We demonstrate a situation that in the absence of dephasing noises the coherent evolution of the system has no role in the energy transport in the network. Therefore, incoherent evolution via application of dephasing noises throughout a selected path of the network leads to transfer the excitation completely to a desired reaction center.
\\
\\
{\bf PACS Nos:}
\end{abstract}

\section{Introduction}
The efficient transport of optical excitation energy through a network of coupled
many-body quantum system has recently become the subject of intense study in both natural and artificial systems \cite{ferrari, ghosh, ghosh2}. Particular examples are energy transport in molecular structure of biological systems ranging their scales from a few atoms to large macro-molecular structures, such as light harvesting (photosynthetic) complexes \cite{Engel, Collini, Hildner}. In general, the overall effect of environment on the quantum transport process in a system is expected to be negative. However, in a large variety of quantum systems, such as chromophoric light-harvesting complexes, the interaction with the environment can result in increased quantum transport efficiency. In fact, interplaying between coherent dynamics and incoherent one gives the optimal way for quantum transport in many noisy systems \cite{garuso, chin, plenio, patrik}. Many efforts have been devoted to study the ways in which quantum transport is optimally affected by the interplay of coherent dynamics and incoherent one arisen from environmental noises \cite{kassal, mohseni, cao, perdomo, vlaming, mohseni2, venuti, hoyer, nalbach, nalbach2, scholak, scholak2, shabani, shabani2}, a phenomenon called environment-assisted quantum transport (ENAQT) or dephasing-assisted quantum transport (DAQT).

On the other hand, there exist quantum systems in which the coherent evolution, due to destructive interferences, is completely suppressed \cite{garuso, chin, plenio}, and therefore, the optimal dynamics of the system is purely incoherent which is issue of  interactions between the system and its fluctuating environments. Destructive interferences can be removed locally or globally in the quantum system when it is affected by its fluctuating environment \cite{patrik}. Indeed, energy transport through pure incoherent evolution in a quantum system can be regarded as a direct evidence for remarkably existence of long-lived quantum coherence and wavelike behavior playing an important role in this way \cite{Engel}.

In this paper, we investigate selective quantum transport of excitation energy throughout a regular network such as a two-dimensional hexagonal-like network of interacting two-level chromophores or sites. As illustrated in the text, there exist a number of sinks considered as reaction centers attached irreversibly to the network. The structure of the network is considered in such way that the coherent part of the dynamics of an excitation created in one of the sites, as initial site, is completely suppressed. Therefore, the selective nature of the transport in the network is related to the selective applications of local independent dephasing noises along a path, as a one-dimensional quantum transport prototype, connecting the initial site to one of the sinks. In fact, application of dephasing noises along a particular path removes the destructive interferences throughout that path in the network. In this way, the excitation energy is transported from the initial site incoherently to the aforementioned sink. This process can also be regarded to implement respectively for the other reaction centers too. On the other hand, to evaluate the optimality of transport in the network, we investigate the optimal effect of dephasing noises on the efficiency of transport along the one-dimensional prototype.  Also, in this way, the effect of energy mismatch between sites on the efficiency of transport is highlighted. It is observed that the optimal transport is robust with respect to the dephasing noises and energy mismatches. Therefore, it is concluded that the optimal conditions for the transport in two-dimensional network lies within the optimal conditions of one-dimensional case.

This paper is organized as follows: In section 2, we demonstrate one-dimensional incoherent quantum transport along with investigation of its optimality conditions which, in turns, are the basic ingredients for quantum transport in two-dimensional hexagonal-like network. Section 3 is devoted for explaining the transport of excitation energy in the two-dimensional network incoherently using local independent dephasing noises. Finally, a brief conclusion is presented in section 4.

\section{One-dimensional quantum transport prototype}

We consider a network as depicted in Fig. 1a, in which the vertices or sites are as two-level chromophoric systems interacting with each others corresponding to edges of the network. The Hamiltonian for this system is considered as
\begin{eqnarray}
&&\hspace{-7mm}{H}=\sum_{j=1}^{3N+1} \hbar\omega_{j}{\sigma}_{j}^{+}{\sigma}_{j}^{-} +\sum_{\{j, l\}\in E}\hbar {\nu}_{j,l} \left({\sigma}_{j}^{-}{\sigma}_{l}^{+}+{\sigma}_{j}^{+}{\sigma}_{l}^{-}\right)
\end{eqnarray}
where $\sigma^{+}_{j}=|j\rangle\langle0|$ and $\sigma^{-}_{j}=|0\rangle\langle j|$
are the raising and lowering operators for a two-level system lied at $j$th vertex or site, the state $|j\rangle$ denotes an excitation in site $j$ and $|0\rangle$ indicates no excitation in that site. The energy of a typical site $j$ is $\hbar\omega_{j}$, and ${\nu}_{j,l}$ is the strength of coupling between $j$th and $l$th sites denotes the hopping rate of an excitation between them and $E$ is the set of edges of the network depicted in Fig. 1a, corresponding to the coupling between the sites.
In general, it is assumed that for the coupling strength we consider $\nu_{3j-2,3j-1}=\nu_{3j-2,3j}$ and $\nu_{3j-1,3j+1}=-\nu_{3j,3j+1}$, and for sites energies $\omega_{3j-1}=\omega_{3j}$ with $j=1, 2, 3, ..., N$. We introduce another set of basis in the single excitation subspace, in terms of standard basis as
\begin{eqnarray}
\begin{array}{c}
  |s_{j}\rangle:=|3j-2\rangle, \quad |s_{j+1}\rangle:=|3j+1\rangle\\\\
  |s^{\pm}_{j}\rangle:=\frac{1}{\sqrt{2}}(|3j-1\rangle\pm|3j\rangle),
\end{array}
\end{eqnarray}
where $j=1, 2, 3, ..., N$. Hence, by choosing the set of basis in (2), the Hamiltonian (1)
is left with a direct sum structure as (see Fig. 1b)
\begin{eqnarray}
&&\hspace{-7mm}H=\bigoplus_{j=1}^{N+1}H_{j},
\end{eqnarray}
where their respective invariant subspaces can be regarded as
\begin{eqnarray}
\begin{array}{c}
  \mathcal{H}_{1}=\mathrm{span}\{|s_{1}\rangle, |s^{+}_{1}\rangle\},\\\\
  \mathcal{H}_{j+1}=\mathrm{span}\{|s^{-}_{j}\rangle, |s_{j+1}\rangle, |s^{+}_{j+1}\rangle\}, j=1, 2, 3, ..., N-1,\\\\
 \mathcal{H}_{N+1}=\mathrm{span}\{|s^{-}_{N}\rangle, |s_{N+1}\rangle\}.
\end{array}
\end{eqnarray}
As it is observed, if at $t=0$, we have an excitation in the initial site of the network ($|1\rangle$) then it is evolved only by the subHamiltonian $H_{1}$ in the invariant subspace $\mathcal{H}_{1}$. Therefore, the excitation can not be received in the site $3N+1$ at all under the pure quantum coherent evolution. This situation is also taken place for the cases in which the excitation at time $t=0$ prepared in another site, because the coherent evolution of the system restricted in only one of the related invariant subspaces. Under these conditions, the evolution of the system can not be performed in the whole of the system. This partial evolution of the system returns, indeed, to the destructive interferences in the network which is a pure quantum effect. Particularly, consider the case in which the site  $3N+1$ of the network has been attached to a reaction center denoted as a sink in such way that the dynamical evolution from the system to the sink is irreversible. Therefore, the population of the sink is always zero when the excitation at $t=0$ is in the initial site of the network ($|1\rangle$).

For the aims of this paper, we consider that $\nu_{3j-2,3j-1}=\nu_{3j-2,3j}=\nu_{3j-1,3j+1}=-\nu_{3j,3j+1}=J$, for $j=1, 2, 3, ..., N$. Now consider a situation in which each of two-level system of the network is in contact with its fluctuating environment. These interactions affect the dynamics of the network in the form of dephasing noise (Fig. 1). Under these considerations, the Lindblad-type master equation for the density matrix of the system is written as
\begin{eqnarray}
&&\hspace{-7mm}\dot{\rho}=\frac{i}{\hbar}[\rho, {H}]+{L}_{deph}(\rho)+{L}_{sink}(\rho)
\end{eqnarray}
where ${L}_{deph}(\rho)$ is the Lindblad operator corresponding to action of the dephasing noises on the $\rho$ given by
\begin{eqnarray}
  {L}_{deph}(\rho)=\sum_{j=1}^{3N+1}(\gamma_{j}(2{\sigma}_{j}^{+}{\sigma}_{j}^{-}\rho{\sigma}_{j}^{+}{\sigma}_{j}^{-}-\{{\sigma}_{j}^{+}{\sigma}_{j}^{-},\rho\}),
\end{eqnarray}
where $\gamma_{1}=\gamma_{2}=. . . =\gamma_{3N+1}=\gamma$, are the rates of dephasing processes that randomize the corresponding phases of local excitations. And also, the additional sink
site is populated by an irreversible decay process from a chosen site (for this case site $3N+1$)
as described by the Lindblad operator
\begin{eqnarray}
&&\hspace{-7mm}{L}_{sink}(\rho)=\Gamma(2{\sigma}_{3N+2}^{+}{\sigma}_{3N+1}^{-}\rho{\sigma}_{3N+1}^{+}{\sigma}_{3N+2}^{-}
-\{{\sigma}_{3N+1}^{+}{\sigma}_{3N+2}^{-}{\sigma}_{3N+2}^{+}{\sigma}_{3N+1}^{-},\rho\}),
\end{eqnarray}
where $\Gamma=2\gamma$ is the rate of dissipative process that reduces the number of excitations in the system and traps it in the sink. The sink population or efficiency of transport is given by
\begin{eqnarray}
P_{sink}(t)=2 \Gamma_{3N+2}\int_{0}^{t}\rho_{_{3N+1,3N+1}}(t')dt'.
\end{eqnarray}
It should be noted that the dynamics preserves the total excitation number in the system and for each $N$ the coherent part of the evolution of the system is completely suppressed. The optimality of incoherent dynamics of the system is defined on the best way of coupling of the system to its independent environments such that the sink site is populated in possibly shortest time. To achieve the optimal conditions for the transport we consider three separate cases. For the first case, it is assumed that all sites of the network with $N=4$ interact, in optimal way, with their respective dephasing environments, Fig. 2a shows the populations of site 1 and the sink versus time. The transfer time for which the sink is completely populated is $t = 505.89$ for this case. The inset in Fig. 2a gives the population of the sink ,$P_{sink}$ , at a fixed time $t = 505.89$, as a function of $\gamma$.

In the second case, the only sites $3j-1$ and $3j$ with $j=1, 2, 3, 4$, are subject to dephasing noises. In optimal way, the efficiency of transport is improved with respect to the previous case (with transfer time $t=391.27$) as shown in Fig. 2b. This shows that the effects of dephasing noises on the sites $3j-2$ and $3j+1$ ($j=1, 2, 3, 4$) reduce the efficiency of transport. In another word, the rate of removing destructive interferences should not be smaller than the rate of decay of phase of the related wave function. For the third case, it is interesting to note that the effect of noises on the only sites $3j-1$ (or $3j$) with j=1, 2, 3, 4, improves the optimal transport better than the previous cases (with transfer time $t=379.1$) as depicted in Fig. 2c. In the next section, we show that when the transport of excitation in a two-dimensional network through an particular one-dimensional path is demanded the conditions in second case is more effective than the others.

Before analyzing the transport process in two-dimensional case, let's consider the robustness of optimal transport versus energy mismatch of sites for the second case. At first, we consider energy of sites $3j-2$ with j=1, 3, 5 as $\hbar(\omega-\delta)$ and with j=2, 4 as $\hbar\omega$. As observed from Fig. 3a, the optimality of transport with respect to $\gamma$ is robust due to the mentioned energy disorders. Now let's consider another case in which for $j=1, 3, 5$ the energy of sites are $\hbar(\omega-\delta)$ and for $j=2, 4$ are $\hbar(\omega+\delta)$ which is more disordered than the first case. As is observed from Fig. 3b, robustness of optimal transport with respect to this type of disordering is less than from the previous one.

\section{Two-dimensional case}
In this stage, we develop the process of excitation transport across a two-dimensional multi-sink network such that the transport can be taken place in completely selective way to each of the reaction centers. Since the evolution is pure incoherent in the network, the transport of an excitation to a particular sink needs to establishing artificial couplings between the network and independent fluctuating environments throughout a path connecting the initial site to that sink. To this end, Let us introduce following Hamiltonian, as a building block for constructing the network, corresponding to Fig. 4, as below
\begin{eqnarray}
\begin{array}{c}
  H_{\mu}=\sum_{j=0}^{3}\hbar \omega_{\mu_{j}}\sigma_{\mu_{j}}^{+}\sigma_{\mu_{j}}^{-}
+\sum_{j=1}^{6}\hbar \omega_{\mu_{0j}}\sigma_{\mu_{0j}}^{+}\sigma_{\mu_{0j}}^{-}+\sum_{j=1}^{6}\hbar \nu_{\mu_{0},\mu_{0j}}(\sigma_{\mu_{0}}^{+}\sigma_{\mu_{0j}}^{-}+\sigma_{\mu_{0}}^{-}\sigma_{\mu_{0j}}^{+}) \\\\
   +\sum_{j=1}^{2}\hbar(\nu_{\mu_{1},\mu_{0j}}(\sigma_{\mu_{1}}^{+}\sigma_{\mu_{0j}}^{-}+\sigma_{\mu_{1}}^{-}\sigma_{\mu_{0j}}^{+})
+\nu_{\mu_{2},\mu_{0j+2}}(\sigma_{\mu_{2}}^{+}\sigma_{\mu_{0j+2}}^{-}+\sigma_{\mu_{2}}^{-}\sigma_{\mu_{0j+2}}^{+})
 \\\\
  +\nu_{\mu_{3},\mu_{0j+4}}(\sigma_{\mu_{3}}^{+}\sigma_{\mu_{0j+4}}^{-}+\sigma_{\mu_{3}}^{-}\sigma_{\mu_{0j+4}}^{+})).
\end{array}
\end{eqnarray}
In general, we assume that $\nu_{\mu_{0},\mu_{01}}=\nu_{\mu_{0},\mu_{02}}$, $\nu_{\mu_{0},\mu_{03}}=\nu_{\mu_{0},\mu_{04}}$, $\nu_{\mu_{0},\mu_{05}}=\nu_{\mu_{0},\mu_{06}}$, $\nu_{\mu_{1},\mu_{01}}=-\nu_{\mu_{1},\mu_{02}}$, $\nu_{\mu_{2},\mu_{03}}=-\nu_{\mu_{2},\mu_{04}}$ and $\nu_{\mu_{3},\mu_{05}}=-\nu_{\mu_{3},\mu_{06}}$ and also $\omega_{\mu_{01}}=\omega_{\mu_{02}}$, $\omega_{\mu_{03}}=\omega_{\mu_{04}}$ and $\omega_{\mu_{05}}=\omega_{\mu_{06}}$. However, for the purpose of this paper, it is enough to rewrite the assumptions as $\nu_{\mu_{0},\mu_{0j}}=J$, $\omega_{\mu_{i}}=\omega_{\mu_{0j}}=\omega$ ($i=0, 1, 2, 4$ and $j=1, 2, . . ., 6$) and $\nu_{\mu_{1},\mu_{01}}=-\nu_{\mu_{1},\mu_{02}}=\nu_{\mu_{2},\mu_{03}}=-\nu_{\mu_{2},\mu_{04}}=\nu_{\mu_{3},\mu_{05}}=-\nu_{\mu_{3},\mu_{06}}=J$. Now we introduce a new set of basis as below
\begin{eqnarray}
\begin{array}{c}
|\mu_{j}\rangle,\quad  j=0, 1, 2, 3, \\\\
|\mu_{1}^{\pm}\rangle=\frac{1}{\sqrt{2}}(|\mu_{01}\rangle\pm|\mu_{02}\rangle),\quad |\mu_{2}^{\pm}\rangle=\frac{1}{\sqrt{2}}(|\mu_{03}\rangle\pm|\mu_{04}\rangle),\quad  |\mu_{3}^{\pm}\rangle=\frac{1}{\sqrt{2}}(|\mu_{05}\rangle\pm|\mu_{06}\rangle).
\end{array}
\end{eqnarray}
Under these considerations, we find that the Hamiltonian $H_{\mu}$, in the new basis, takes a direct sum structure as
\begin{eqnarray}
  H_{\mu}=\bigoplus_{j=0}^{3}H_{\mu_{j}},
\end{eqnarray}
where their corresponding representative subspaces are
\begin{eqnarray}
\begin{array}{c}
  \mathcal{H}_{\mu_{0}}=\mathrm{span}\{|\mu_{0}\rangle, |\mu_{1}^{+}\rangle, |\mu_{2}^{+}\rangle, |\mu_{3}^{+}\rangle\},\\\\
 \mathcal{H}_{\mu_{j}}=\mathrm{span}\{|\mu_{j}\rangle, |\mu_{j}^{-}\rangle\},\quad  j=1, 2, 3.
 \end{array}
\end{eqnarray}
As the one-dimensional quantum transport prototype, evolution of the closed system described by the Hamiltonian $H_{\mu}$ is governed by one of subHamiltonian of Eq. 11 and contained in one of the respective invariant subspaces of Eq. 12 (depends on the position of initial state). If an excitation is prepared at site $\mu_{1}$ as an initial state then it can not be transferred to site $\mu_{2}$ nor $\mu_{3}$ because they lie in different invariant subspaces. However, if some of these invariant subspaces are tailored to each other by some processes such as interacting the network at sites, for example, $\mu_{0j}$ with $j=1, 2, 3, 4$ ($j=1, 2, 5, 6$) with fluctuating environments then the  quantum transport of excitation can be possible from $\mu_{1}$ to $\mu_{2}$ ($\mu_{3}$), as seen in Figs. 4.

Now, using the last discussion and Figs. 4 as a building block, a two-dimensional quantum network can be constructed whereby energy of excitation can be transferred from an initial site to one of reaction centers attached to the network. Consider, for example, a network with three identical sinks as reaction centers corresponding to the Fig. 5, in which the complete transport of excitation prepared in site 1 to arbitrarily one of the reaction centers is demanded. The Hamiltonian of the system is given as
\begin{eqnarray}
 H=\sum_{\mu}H_{\mu},
\end{eqnarray}
where $H_{\mu}$ is as in Eq. 9. It is clear that the Hamiltonian of the network has also direct sum structure therefore, evolution of the system restricted to one the related invariant subspaces. The direct sum structure of $H$ is as \begin{eqnarray}
\begin{array}{c}
  H=H_{\mu_{1}}\oplus H_{\lambda_{2}}\oplus H_{\nu_{3}}, \\\\
  \qquad\oplus H_{\mu_{0}}\oplus H_{\lambda_{0}}\oplus H_{\nu_{0}}, \\\\
  \qquad\oplus H_{\mu,\lambda}\oplus H_{\mu,\nu}\oplus H_{\lambda,\nu},
\end{array}
\end{eqnarray}
 and their corresponding invariant subspaces are
 \begin{eqnarray}
\begin{array}{c}
  \mathcal{H}_{a}=\mathrm{span}\{|a\rangle, |a^{-}\rangle\}, \quad a=\mu_{1}, \lambda_{2}, \nu_{3}, \\\\
  \mathcal{H}_{b0}=\mathrm{span}\{|b_{0}\rangle, |b_{1}^{+}\rangle, |b_{2}^{+}\rangle, |b_{3}^{+}\rangle\}, \quad b=\mu, \lambda, \nu, \\\\
  \mathcal{H}_{\mu,\lambda}=\mathrm{span}\{|\mu_{2}^{-}\rangle, |\mu_{2}(\lambda_{1})\rangle, |\lambda_{1}^{-}\rangle\},\\\\
  \mathcal{H}_{\mu,\nu}=\mathrm{span}\{|\mu_{3}^{-}\rangle, |\mu_{3}(\nu_{1})\rangle, |\nu_{1}^{-}\rangle\},\\\\
  \mathcal{H}_{\lambda,\nu}=\mathrm{span}\{|\lambda_{3}^{-}\rangle, |\lambda_{3}(\nu_{2})\rangle. |\nu_{2}^{-}\rangle\},
\end{array}
\end{eqnarray}
As discussed previously, evolution of an excitation prepared at site 1 is only restricted within the invariant subspace $H_{\mu_{1}}$ (Fig. 5). Consider that the sinks are attached to the sites $\lambda_{2}$, $\lambda_{3}$ (or $\nu_{2}$) and $\nu_{3}$ with equal strength of coupling as $\Gamma=2\gamma$ where $\gamma$ is the rate of dephasing noise on a typical site. The populations of sinks are
\begin{eqnarray}
\begin{array}{c}
             P_{sink 1}(t)=2 \Gamma\int_{0}^{t}\rho_{_{\lambda_{2},\lambda_{2}}}(t')dt' ,\\\\
             P_{sink 2}(t)=2 \Gamma\int_{0}^{t}\rho_{_{\lambda_{3},\lambda_{3}}}(t')dt' ,\\\\
             P_{sink 3}(t)=2 \Gamma\int_{0}^{t}\rho_{_{\nu_{3},\nu_{3}}}(t')dt' .
           \end{array}
\end{eqnarray}

Now consider the effect of dephasing noises with equal rate $\gamma$ on the sites $\mu_{01}$, $\mu_{02}$, $\mu_{03}$, $\mu_{04}$, $\lambda_{01}$, $\lambda_{02}$, $\lambda_{03}$, $\lambda_{04}$. We see that the excitation is transferred optimally from site 1 (or $\mu_{1}$) to the sink 1 completely without any penetration to the other sinks as shown in Fig. 6a. On the other hand, the excitation completely transfers to the sink 2 if the dephasing noises affect the sites $\mu_{01}$, $\mu_{02}$, $\mu_{03}$, $\mu_{04}$, $\lambda_{01}$, $\lambda_{02}$, $\lambda_{05}$, $\lambda_{06}$, as is obvious from Fig. 6b. And in similar way, if the sites $\mu_{01}$, $\mu_{02}$, $\mu_{05}$, $\mu_{06}$, $\nu_{01}$, $\nu_{02}$, $\nu_{05}$, $\nu_{06}$ are affected by dephasing noises sink 3 is only populated (Fig. 6c). Hence, transport of excitation to one of reaction centers is possible only through incoherent coupling of the network to dephasing environments, which in turn, tailor a number of invariant subspaces throughout the path connecting the site 1 to the desired reaction center. The transport process discussed here for each sink through a particular one-dimensional path is optimal and similar to the conditions of second case of one-dimensional prototype discussed in previous section.

The discussed scheme for selective transport of energy can be extended for another larger networks easily (Fig. 7). In general, the conditions for the respective optimal transport will be different.

\section{Conclusions}

In this work, we have presented a method for energy transport in a two-dimensional network in a selective way. In this approach coherent part of the evolution is an unwanted process for selective transfer of excitation energy. Therefore, the network is designed in such way that the coherent evolution is completely suppressed by itself. So the evolution of the system is completely incoherent whose existence depends on the existence of interactions between the system and  independent environmental fluctuating noises. If the interactions are established throughout a particular path of the network the evolution takes place along that path incoherently. Specially, the path can be considered as a connection link between the site 1 (the excitation was prepared initially in this site) and one of sinks or reaction centers, so in this way, the excitation can be transferred completely to the respective reaction center. On the other hand, it was observed that the optimal transport throughout a particular path in the two-dimensional network is not similar to the optimality of quantum transport for one-dimensional prototype. It is interesting to note that since the evolution takes place in a particular path rather than in whole of the network so, from the dissipation and losing point of view which can be occurred in chromophores or sites, the quantum transport can be performed in more efficient way which can be led to further investigations in future.

An additional point of view is the effect of reorganization energy shift for each site which has interaction with the related environment. Since all of the interacting sites with the environments are identical (so are the environments), each of them experiences identical energy shift. Optimality of transport under this kind of energy shifts, for Markovian and non-Markovian environments, can also be investigated in future.

\newpage
Fig. 1. (a) A network of two-level systems coupled to each other as $\nu_{3j-2, 3j-1}=\nu_{3j-2, 3j}=\nu_{3j-1, 3j+1}=-\nu_{3j, 3j+1}=J$ and attached to independent dephasing noises. (b) The same network under the change of basis. Invariant subspaces are connected incoherently by the dephasing noises.
\begin{figure}
\centering
\includegraphics[width=445 pt]{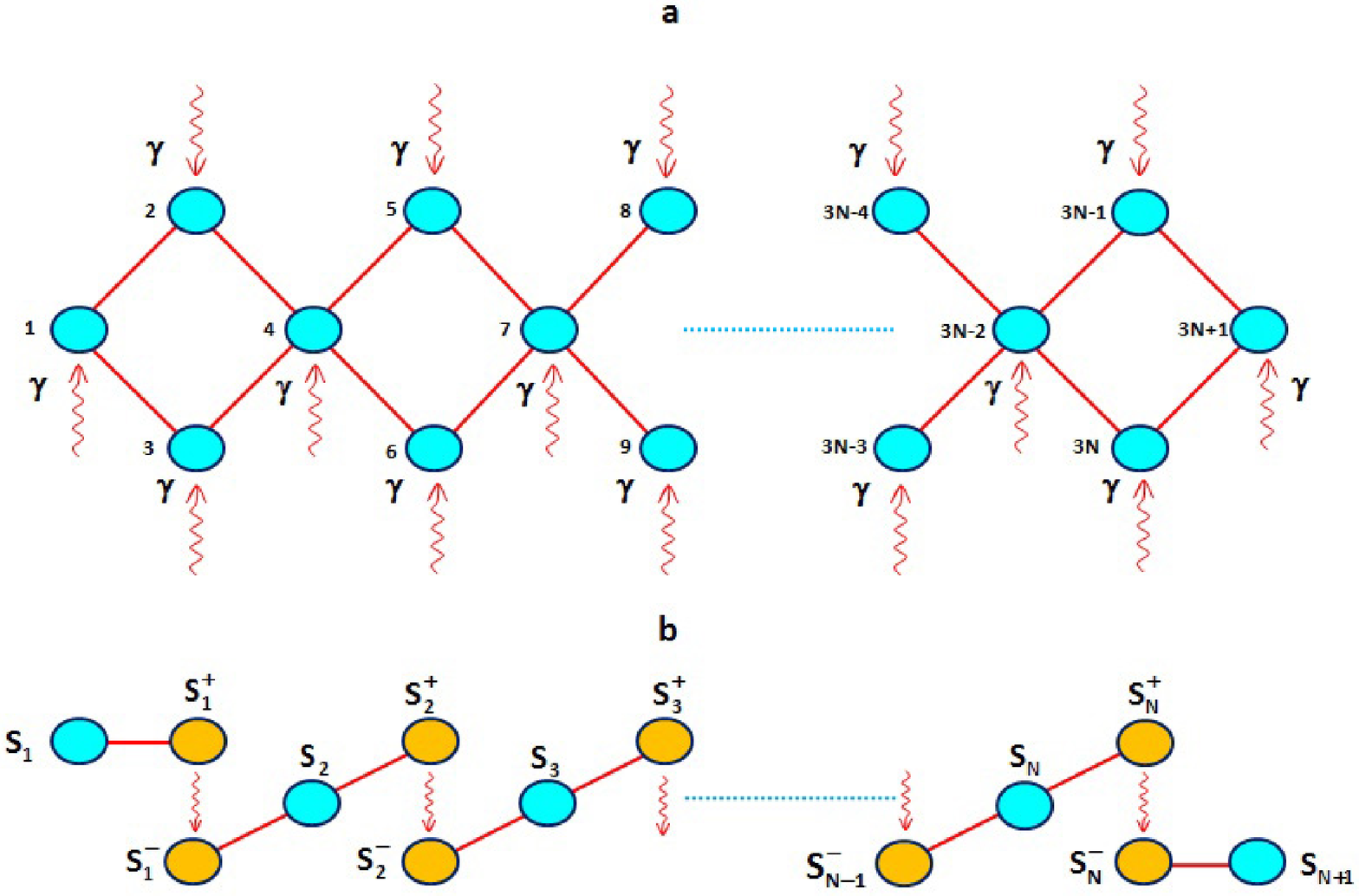}
\caption{} \label{Fig2}
\end{figure}
\newpage
Fig. 2. Populations of site 1 and the sink for $N=4$ corresponding to the Fig. 1, with $\omega_{1}=\omega_{2}=... =\omega_{13}=50$. The network is affected by independent dephasing noises in three different ways: (a) all sites are attached to the noise with optimal rate $\gamma_{Opt}=0.95$, (b) the sites $3j-1$ and $3j$, ($j=1, 2, 3, 4$), are attached with $\gamma_{Opt}=1.22$ and (c) the sites $3j-1$, ($j=1, 2, 3, 4$), are attached with $\gamma_{Opt}=2.44$. Each inset shows the dependence of $P_{sink}$ at a fixed time (the respective transfer time) as a function of $\gamma$. The initially sharp rise is due to the increasing rapidity of destruction of invariant subspaces while the decreasing rate is due to the quantum Zeno effects.
\begin{figure}
\centering
a\\
    \centering
        {
        \includegraphics[width=3.5in]{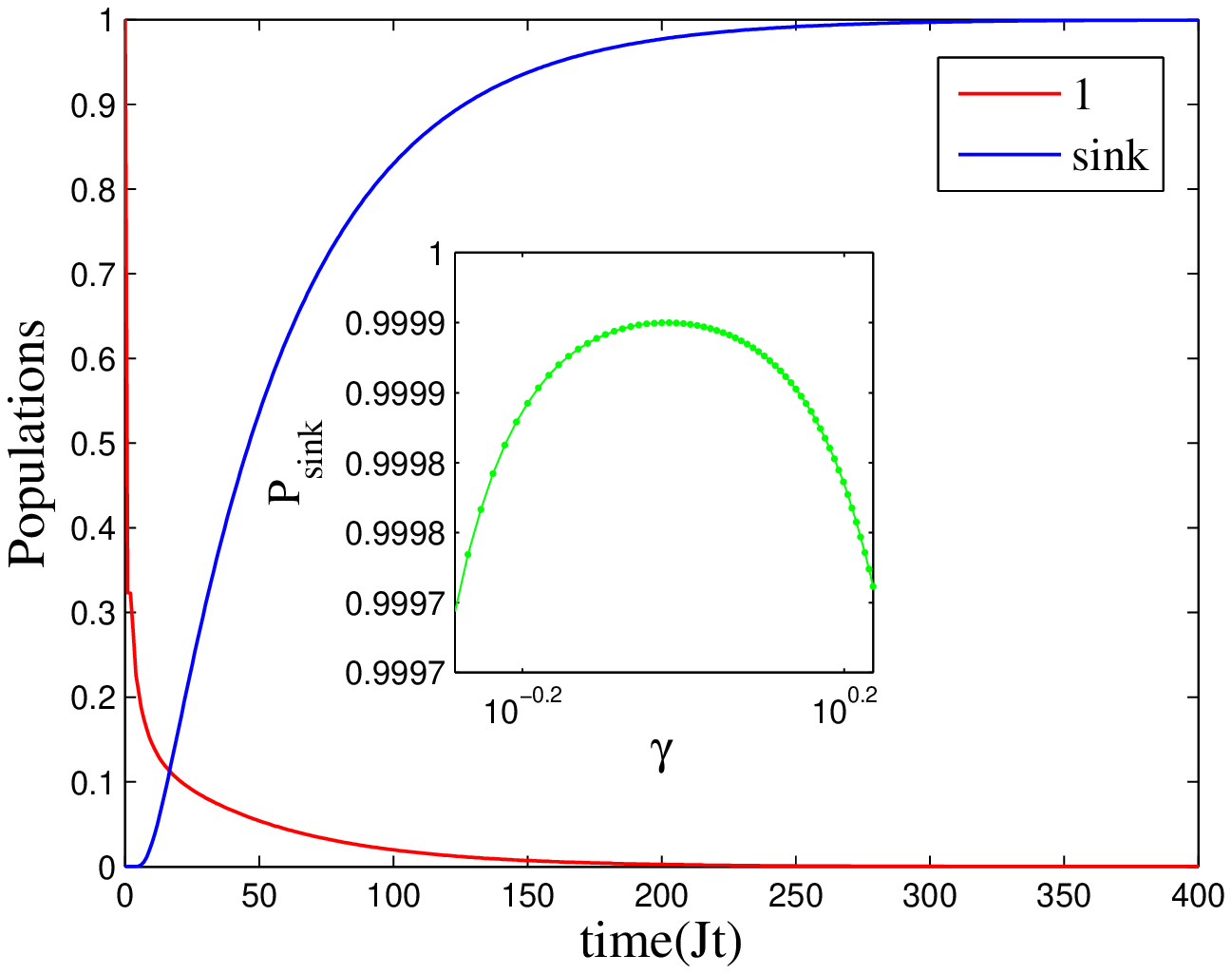}
        \label{fig:first_sub}
    }
    \\
    \centering
b\\
    \centering
       {
        \includegraphics[width=3.5in]{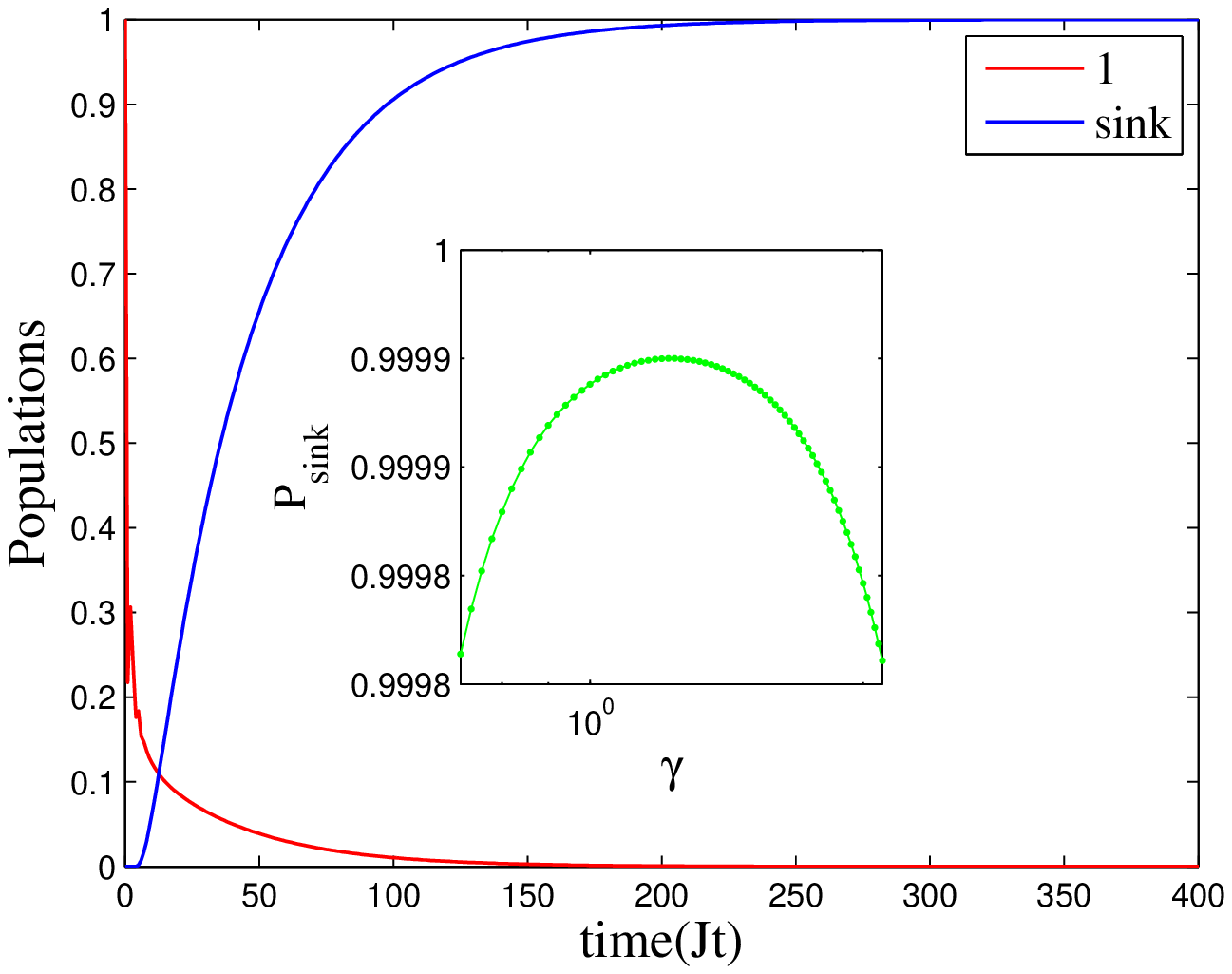}
        \label{fig:second_sub}
    }\\
    \centering
c\\
    \centering
        {
        \includegraphics[width=3.5in]{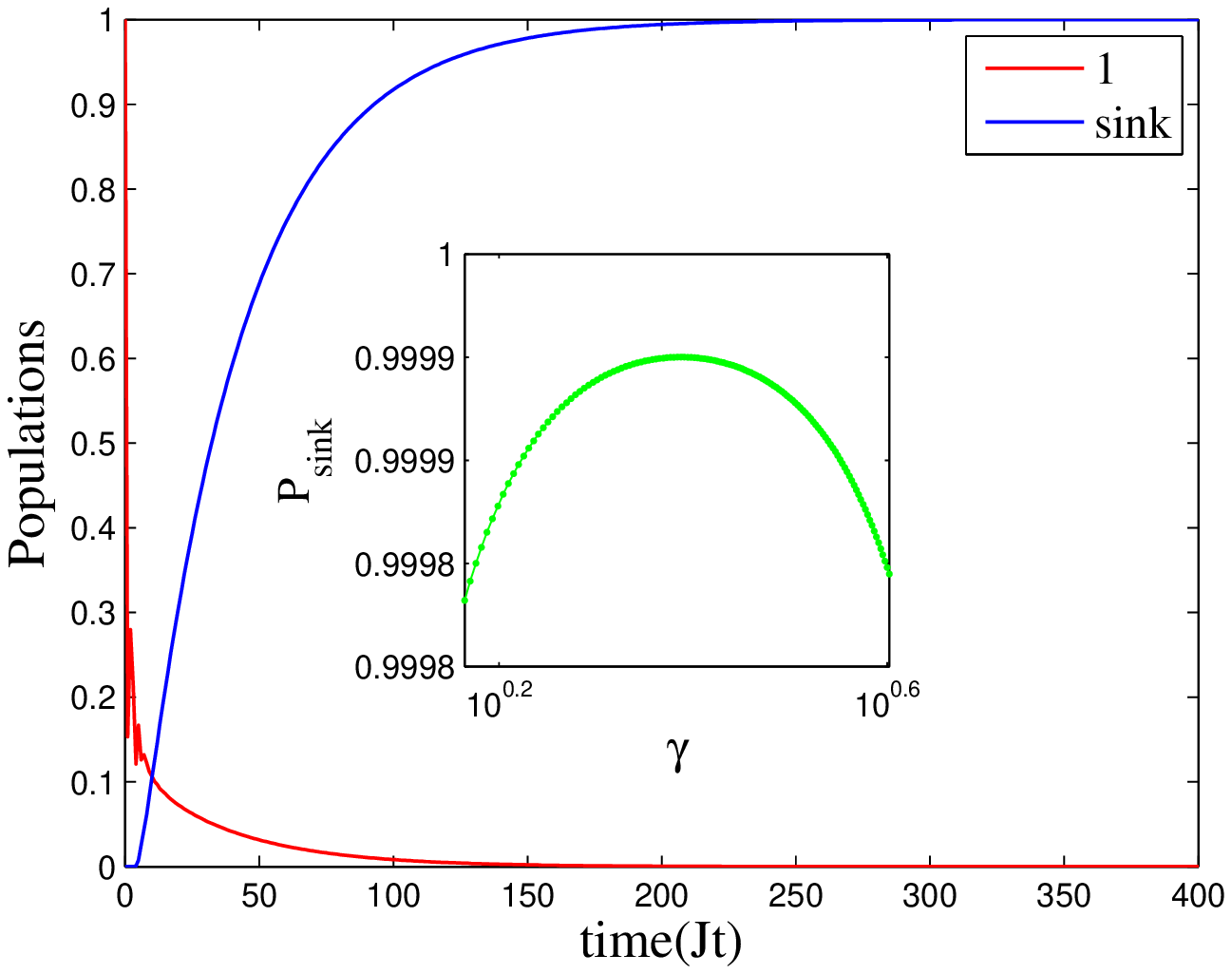}
        \label{fig:third_sub}
    }
    \centering
    \caption{}
\end{figure}
\newpage
Fig. 3. Robustness of $P_{sink}$ around the optimal value of dephasing noise due to energy mismatch between sites. (a) The energies of sites $3j-2$ with j=1, 3, 5 are $\hbar(\omega-\delta)$ and with j=2, 4 are $\hbar\omega$. (b) The energies of sites $3j-2$ with j=1, 3, 5 are $\hbar(\omega-\delta)$ and with j=2, 4 are $\hbar(\omega+\delta)$. For both cases the sites $3j-1$ and $3j$ are attached to dephasing noises.
\begin{figure}
\qquad\qquad\qquad\qquad a \qquad\qquad\qquad\qquad\qquad\qquad\qquad\qquad\qquad\qquad b \\
        {
        \includegraphics[width=3.3in]{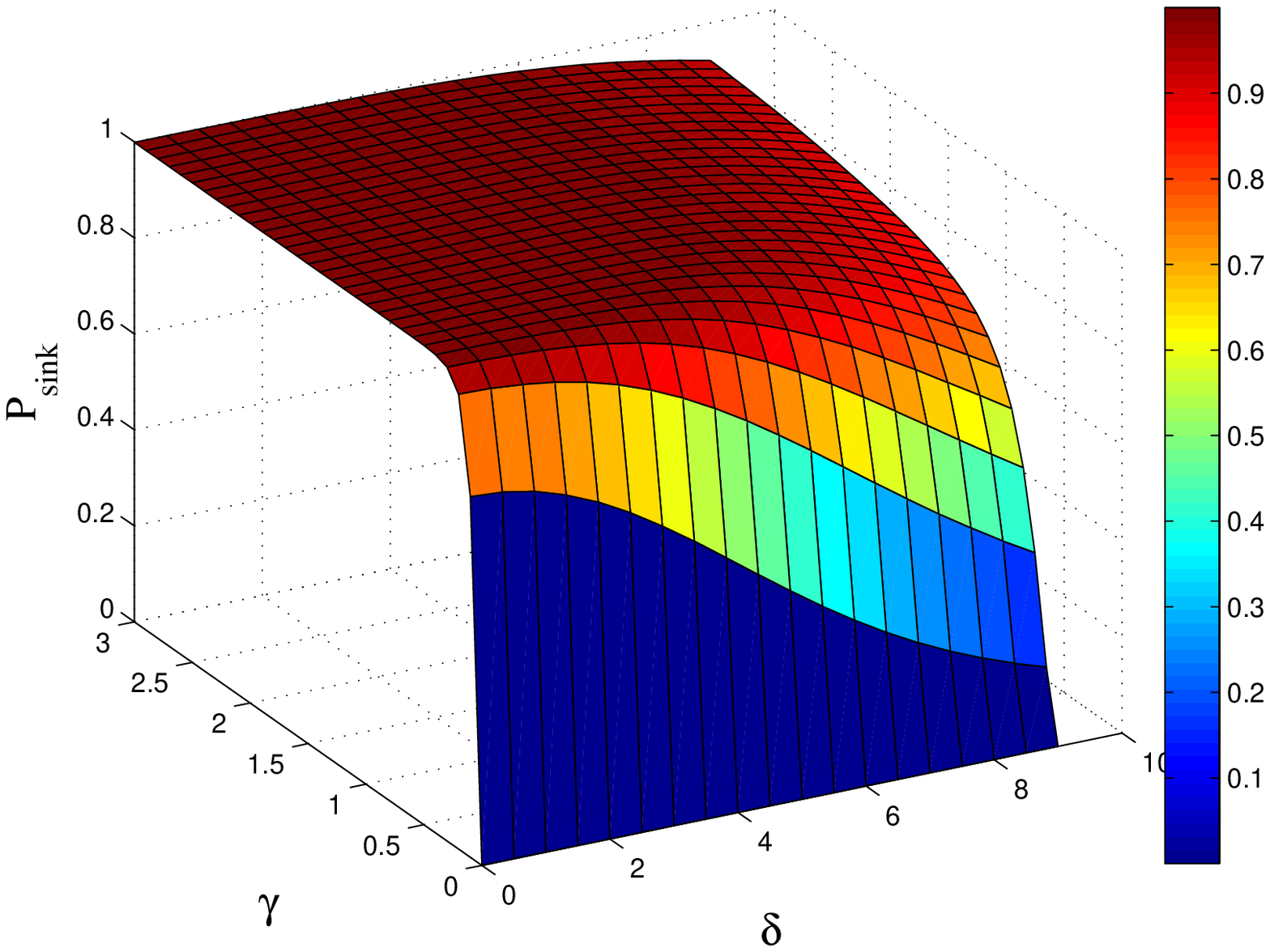}
        \label{fig:first_sub}
    }{
        \includegraphics[width=3.3in]{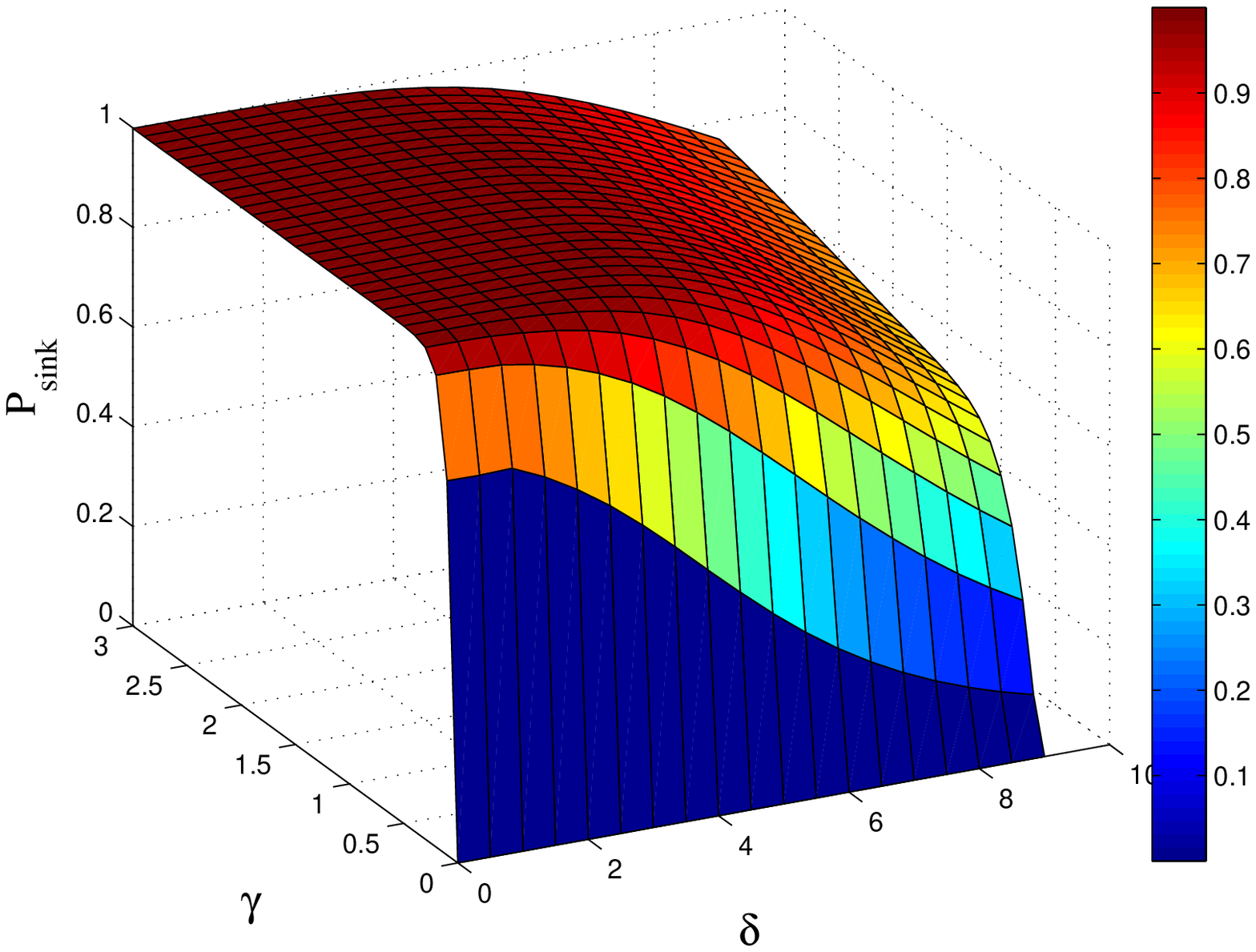}
        \label{fig:second_sub}
    }\\
        \caption{}
\end{figure}
\newpage
Fig. 4. Selective incoherent quantum transport building block in two dimension. (a) Incoherent transport from site $\mu_{1}$ to site $\mu_{2}$ and (b) incoherent transport from site $\mu_{1}$ to site $\mu_{3}$. (c) and (d) show invariant subspaces structure of (a) and (b) respectively, and incoherent connection between them through the noises.
\begin{figure}
        \qquad \qquad\qquad\qquad \qquad a \qquad\qquad \qquad\quad\qquad\qquad\qquad\qquad\quad\qquad\qquad\qquad b\\{
        \includegraphics[width=3.6in]{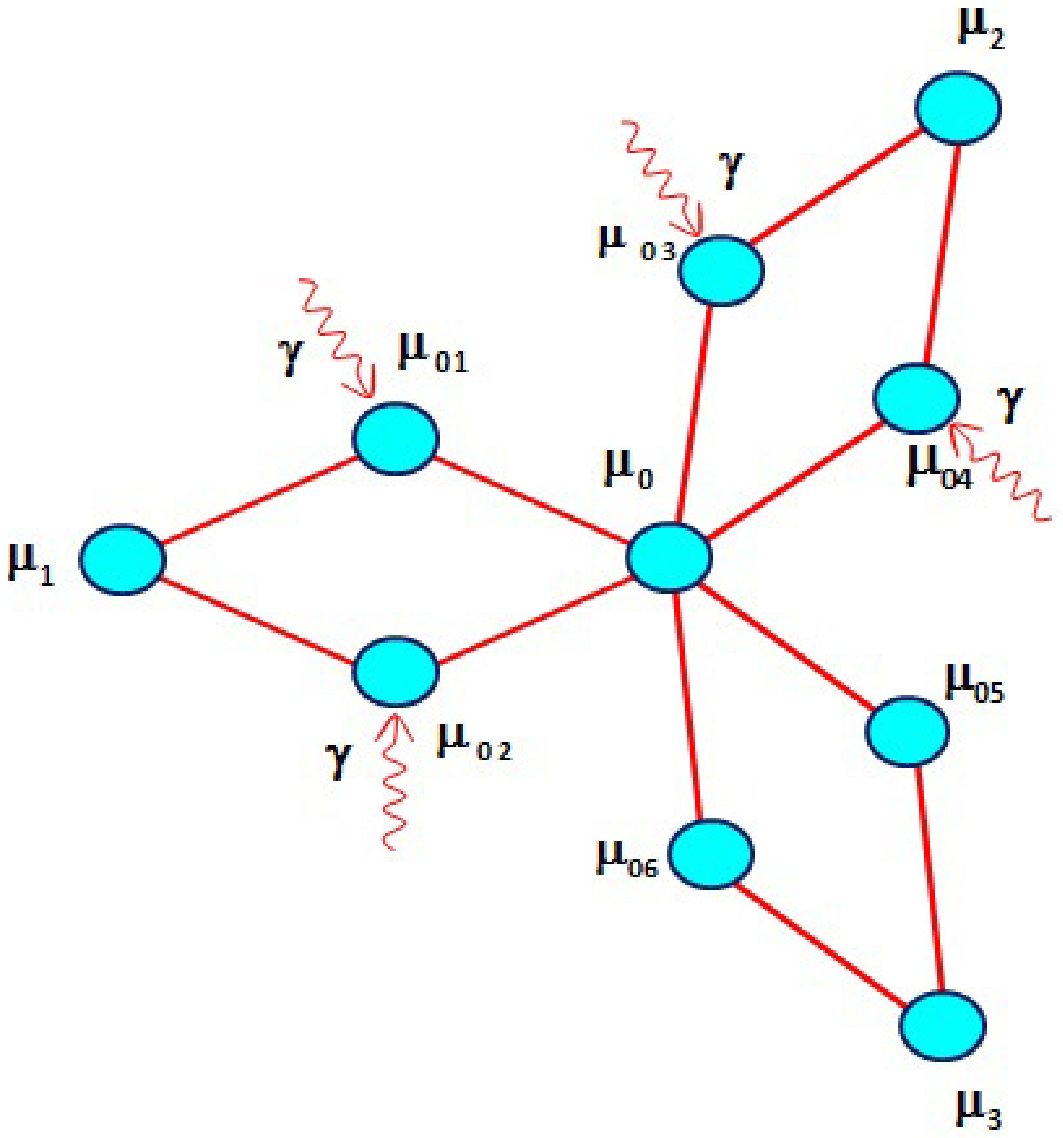}
        \label{fig:first_sub}
    }{
        \includegraphics[width=3.6in]{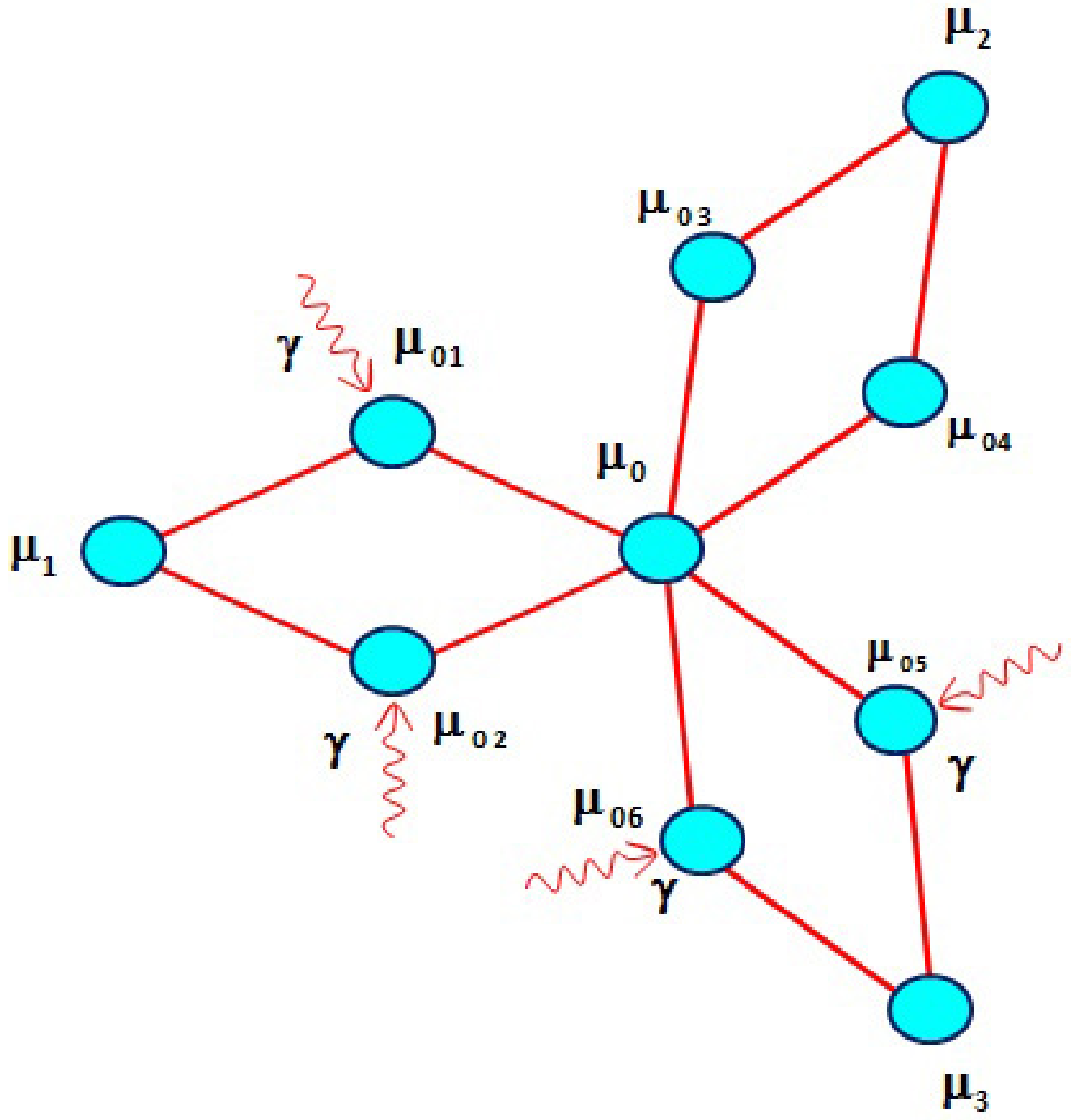}
        \label{fig:second_sub}
    }\\ \par \quad \quad\qquad\qquad\qquad \qquad c \qquad\qquad \qquad\quad\qquad\qquad\qquad\qquad\quad\qquad\qquad\qquad d\\{
        \includegraphics[width=3.5in]{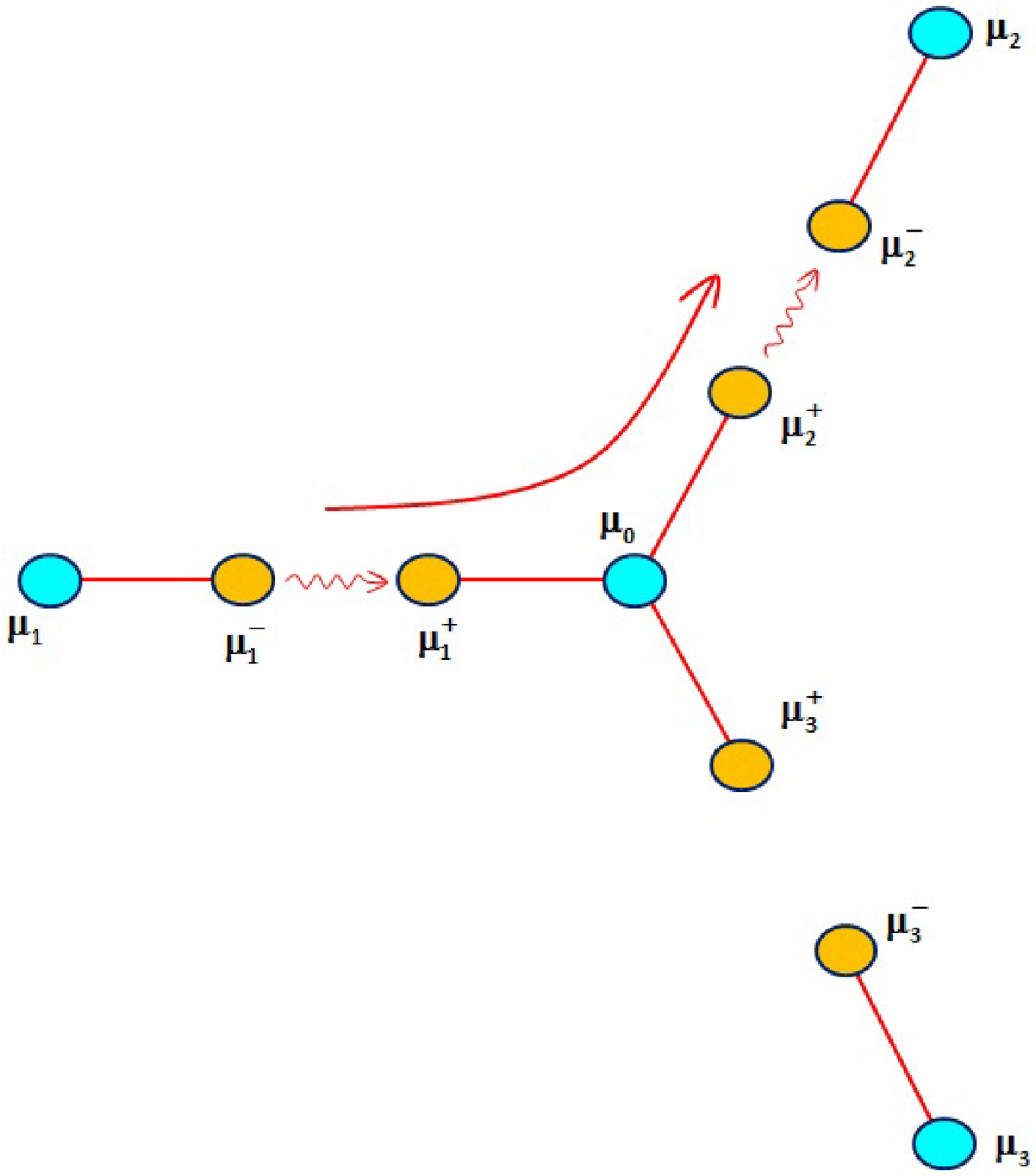}
        \label{fig:first_sub}
    }{
        \includegraphics[width=3.5in]{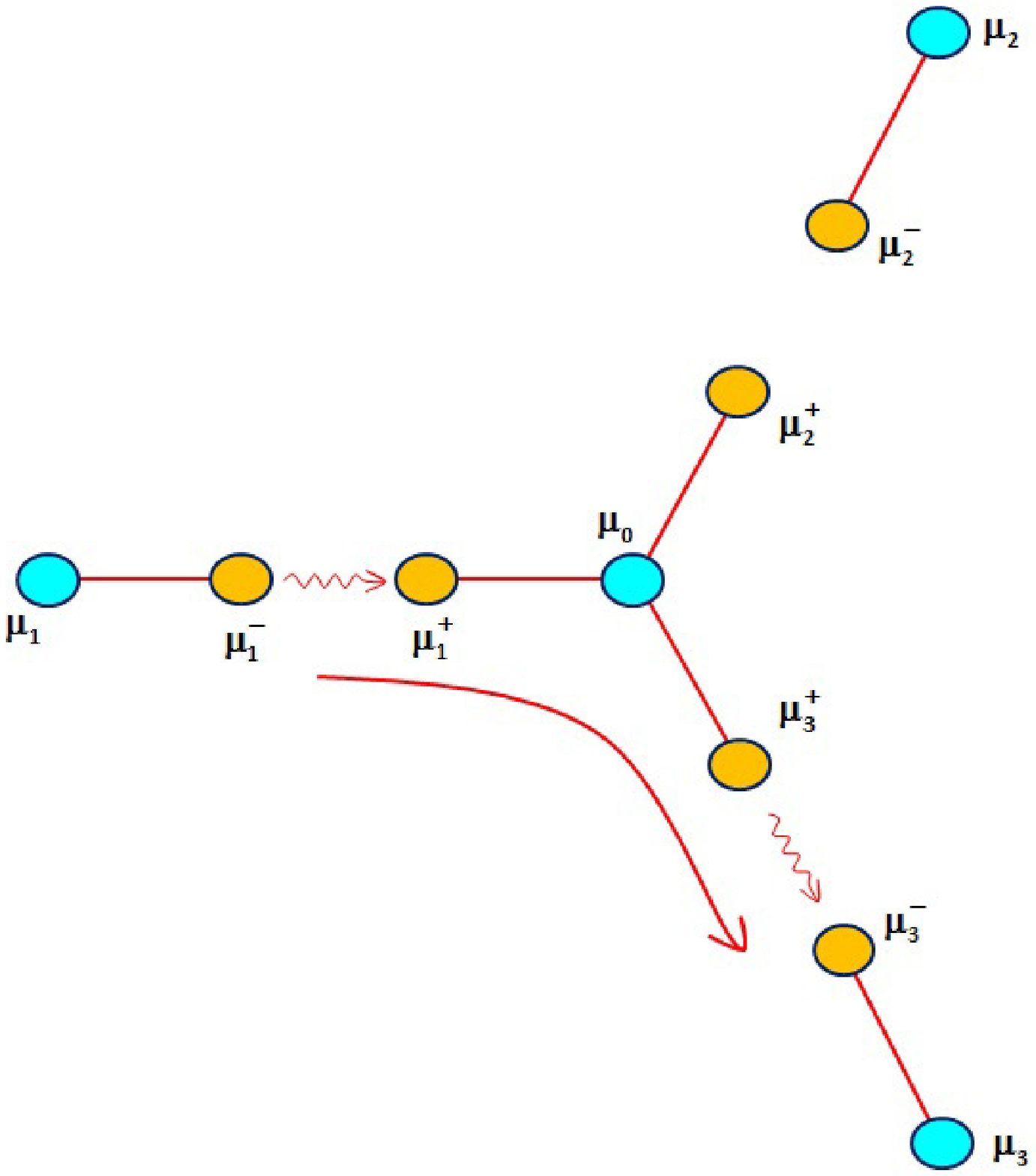}
        \label{fig:second_sub}
    }
    \caption{}
    \end{figure}
\newpage
Fig. 5. A network with three sinks. Energy excitation can be transferred from site 1 to one of the sinks arbitrarily.
\begin{figure}
\centering
\includegraphics[width=445 pt]{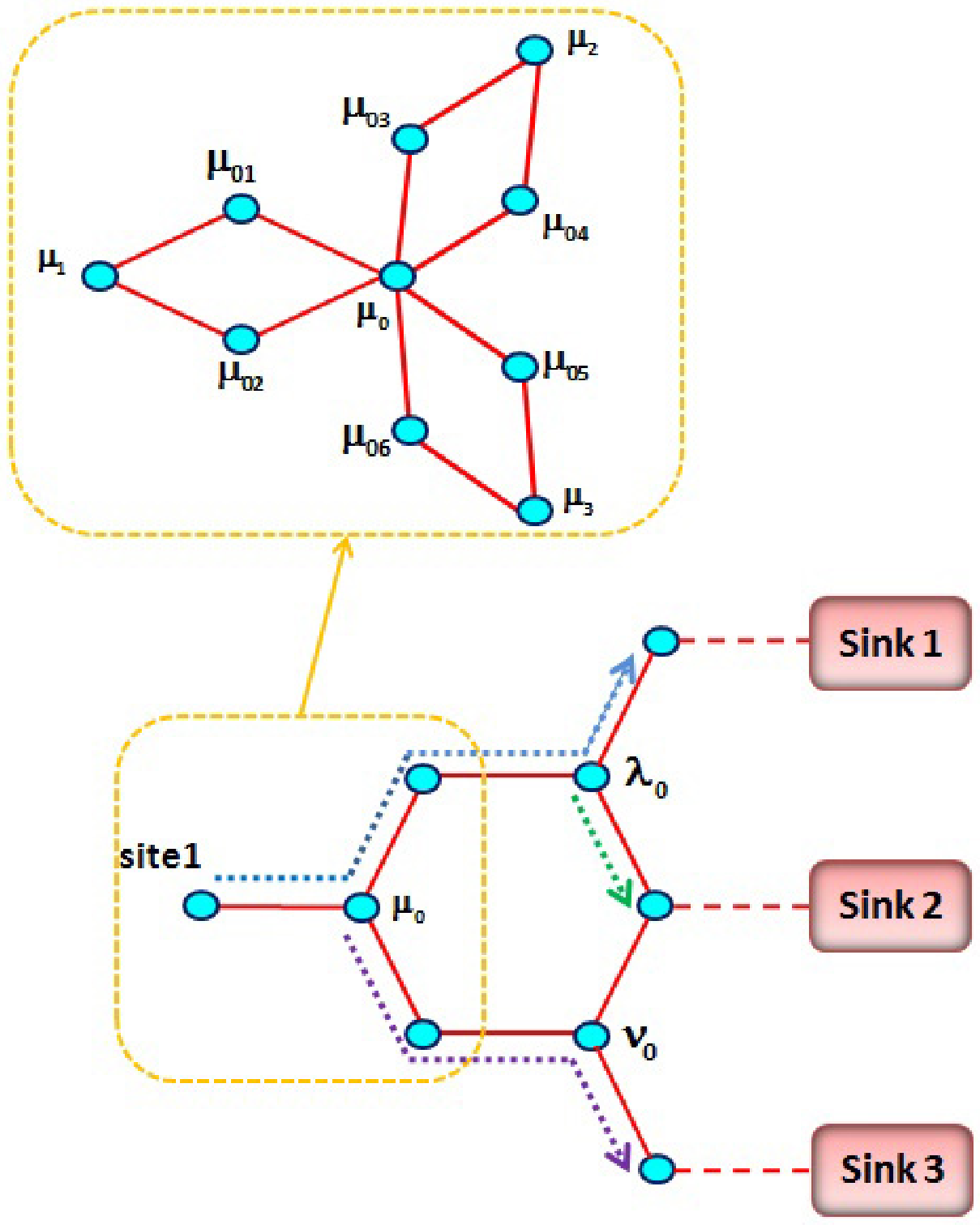}
\caption{}
\end{figure}
\newpage
Fig. 6. Complete transfer of excitation from site 1, selectively to (a) sink 1, (b) sink 2 and (c) sink 3.
\begin{figure}
\centering
a\\
    \centering
        {
        \includegraphics[width=3.7in]{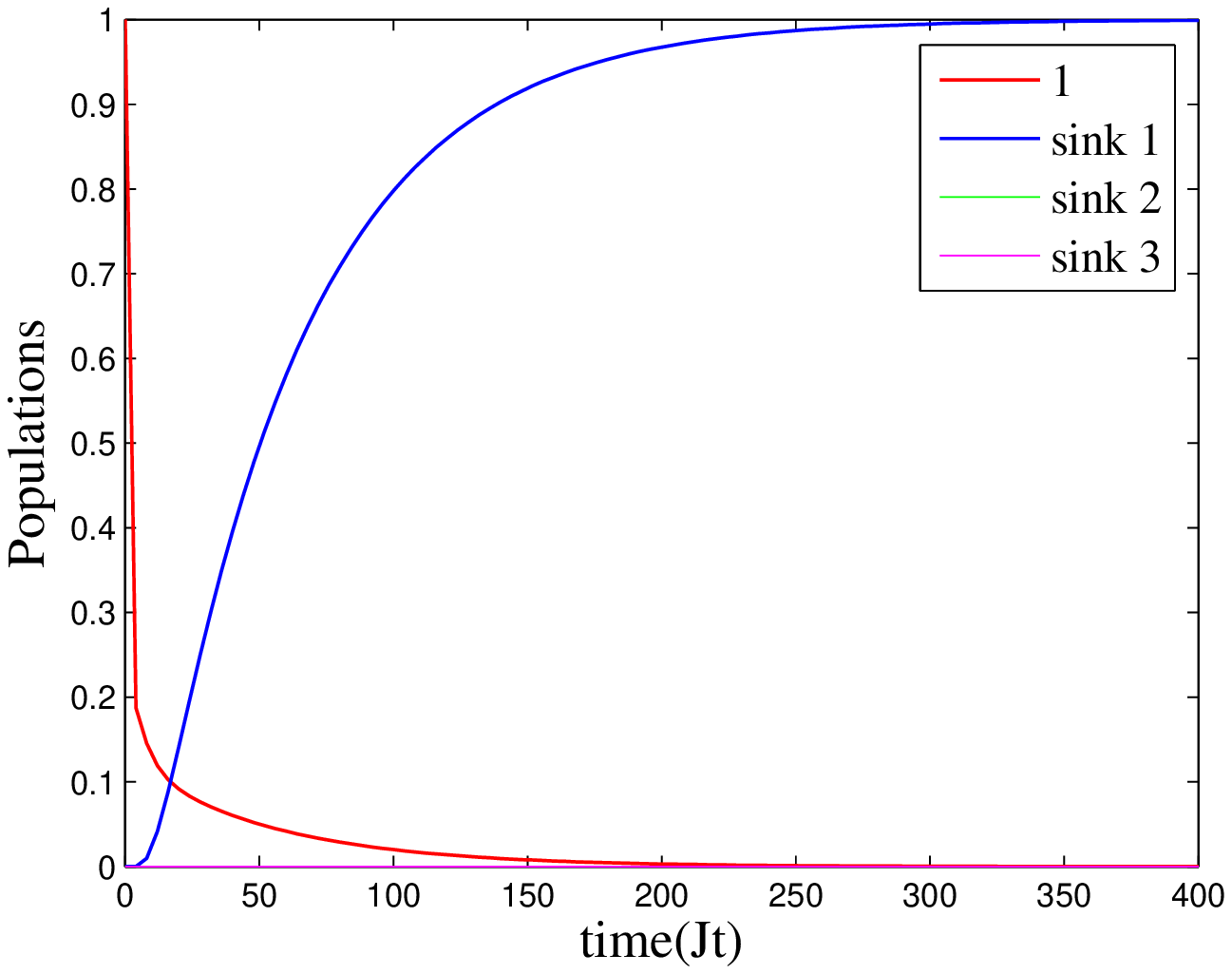}
        \label{fig:first_sub}
    }
    \\
\centering
b\\
    \centering
       {
        \includegraphics[width=3.7in]{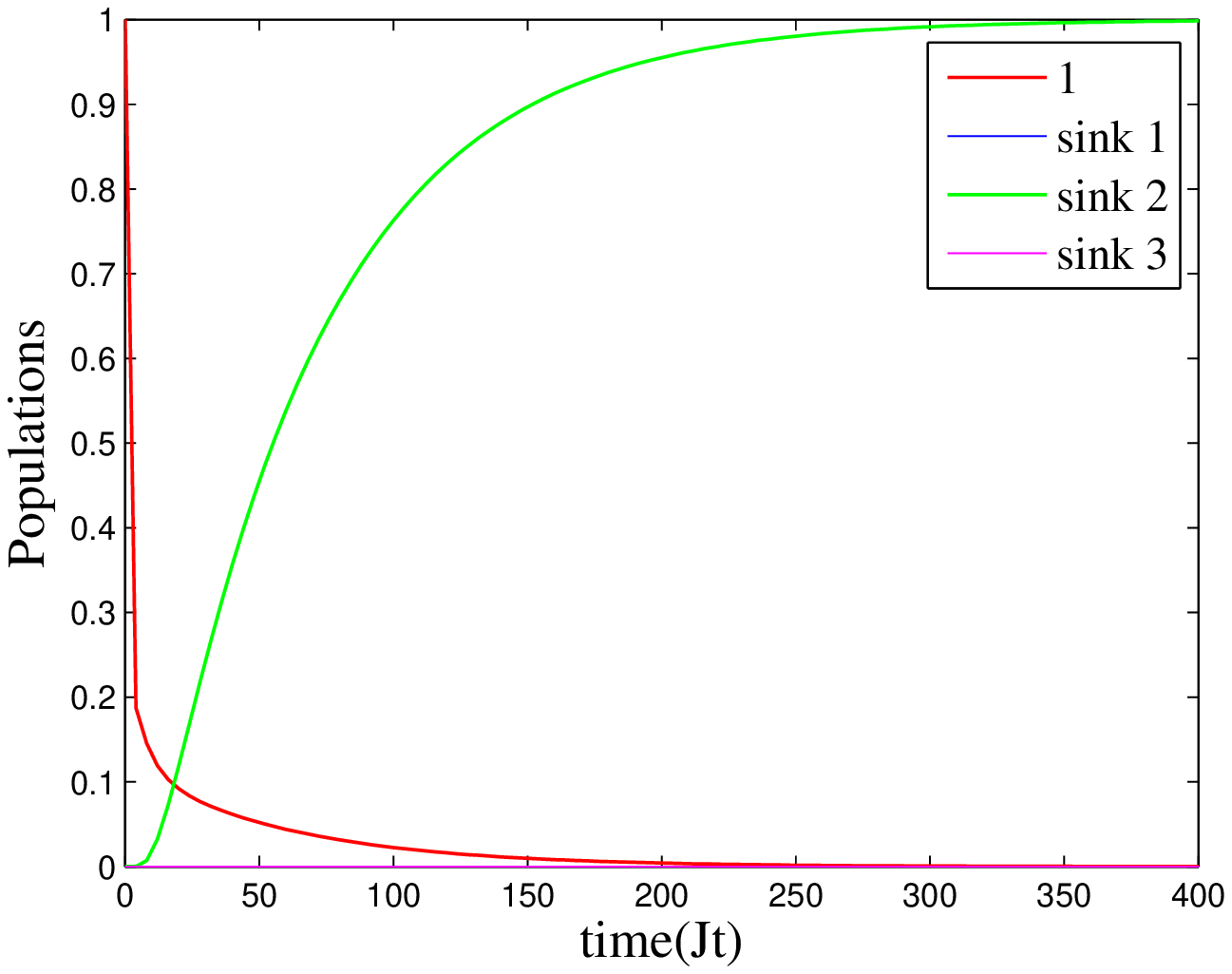}
        \label{fig:second_sub}
    }\\
\centering
c\\
    \centering
        {
        \includegraphics[width=3.7in]{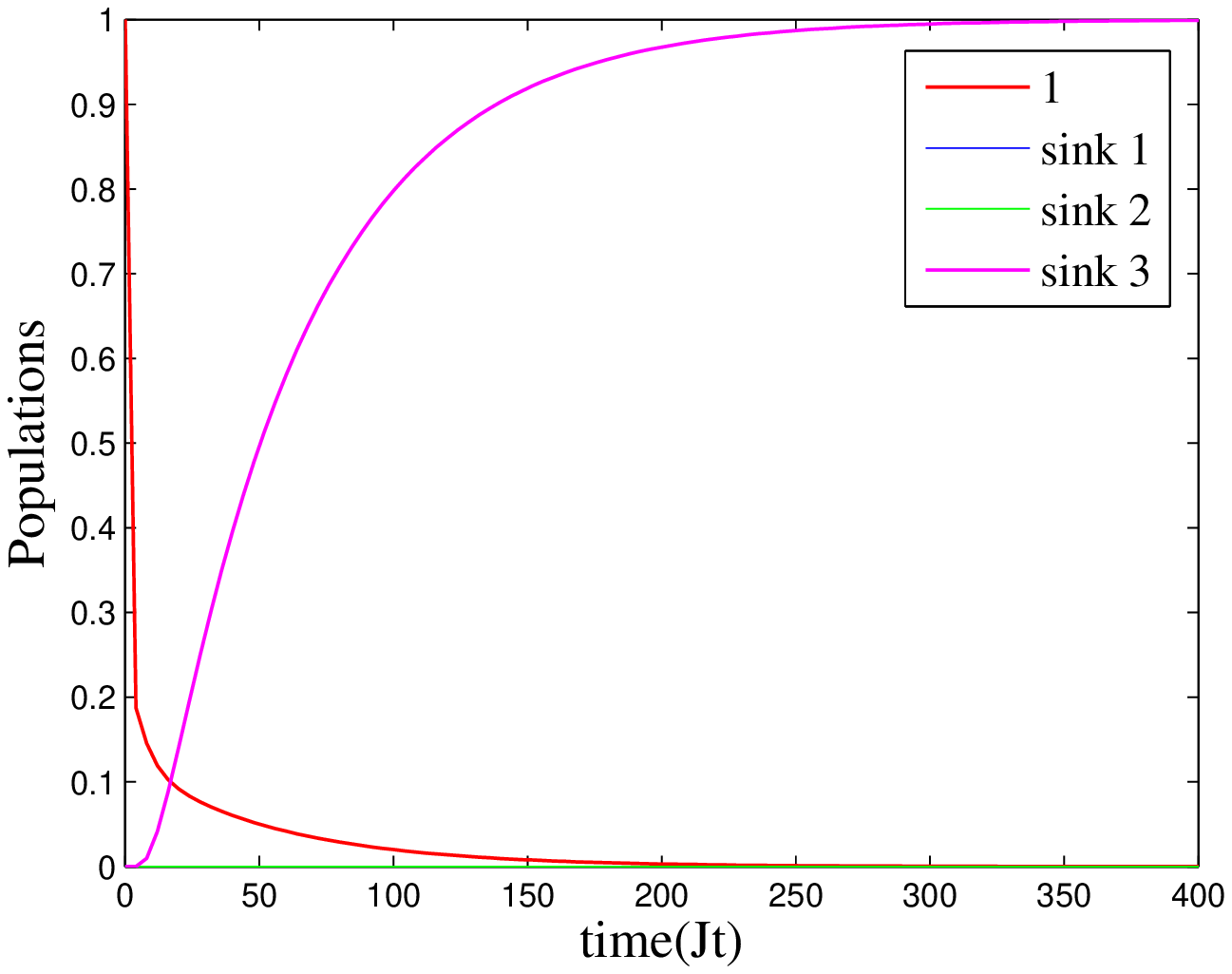}
        \label{fig:third_sub}
    }
    \caption{}
    \end{figure}
\newpage
Fig. 7. Extended network with more than three sinks.
\begin{figure}
\centering
\includegraphics[width=445 pt]{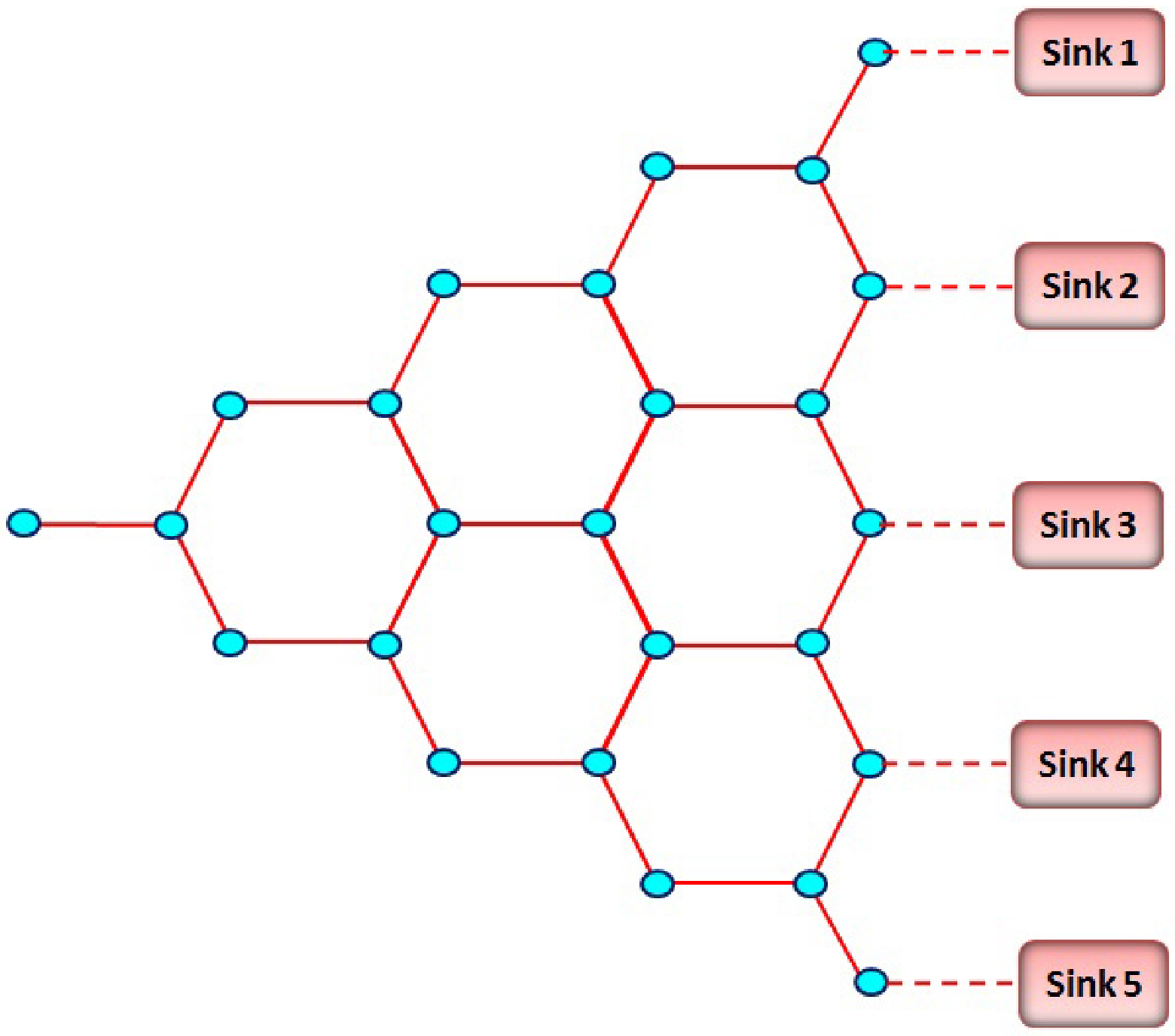}
\caption{}
\end{figure}
\end{document}